\documentclass{article}
\usepackage{INTERSPEECH2022,amsmath,graphicx,acronym,balance}
\usepackage{amssymb,amsmath} 
\usepackage{url}
\usepackage[subtle]{savetrees} 
\usepackage{cite}
\usepackage[table,dvipsnames]{xcolor}
\usepackage{enumitem}
\usepackage{multirow}
\usepackage{graphics}
\usepackage{paralist}
\usepackage{xcolor}
\usepackage{soul}

\setdefaultleftmargin{1em}{2em}{}{}{}{}
\setitemize{noitemsep,topsep=0pt,parsep=0pt,partopsep=0pt}

\acrodef{STFT}{short-time Fourier transform}
\acrodef{iSTFT}{inverse short-time Fourier transform}
\acrodef{RIR}{room impulse response}
\acrodef{RUNet}{recurrent convolutional network with U structure} 
\acrodef{CRN}{convolutional recurrent network}
\acrodef{FC}{fully connected}
\acrodef{TRUNet}{transformer-recurrent-U network for multi-channel reverberant sound source separation}

\title{TRUNet: Transformer-Recurrent-U Network for End-to-end\\ Multi-channel Reverberant Sound Source Separation}

\name{%
    Ali Aroudi, Stefan Uhlich, Marc Ferras Font
    }
\address{%
    Sony Europe B.V., Stuttgart, Germany 
        } 
\email{}

\begin{document}
\ninept
\maketitle

\begin{abstract}

\end{abstract}
In recent years, many deep learning techniques for single-channel sound source separation have been proposed using recurrent, convolutional and transformer networks. When multiple microphones are available, spatial diversity between speakers and background noise in addition to  spectro-temporal diversity can be exploited by using multi-channel filters for sound source separation. Aiming at end-to-end multi-channel source separation, in this paper we propose a transformer-recurrent-U network (TRUNet), which directly estimates multi-channel filters from multi-channel input spectra. TRUNet consists of a spatial processing network with an attention mechanism across  microphone channels aiming at capturing the spatial diversity, and a spectro-temporal processing network aiming at capturing spectral and temporal diversities. In addition to multi-channel filters, we also consider estimating single-channel filters from multi-channel input spectra using TRUNet. We train the network on a large reverberant dataset using a proposed combined compressed mean-squared error loss function, which further improves the sound separation performance. We evaluate the network on a realistic and challenging reverberant dataset, generated from measured room impulse responses of an actual microphone array. The experimental results on realistic reverberant sound source separation show that the proposed TRUNet outperforms state-of-the-art single-channel and multi-channel source separation methods.


%
\section{Introduction}
\label{sec:introduction}

Speech signals captured by microphones placed at a distance from speakers are often corrupted with various undesired acoustic sources, such as interfering speakers, reverberation and ambient noise, which lead to a decreased speech quality and intelligibility. 
In recent years, aiming at separating out the speakers from the microphone signals and reduce background noise, sound source separation techniques based on deep learning have been proposed. Sound source separation techniques can be broadly categorized into single-channel and multi-channel methods, based on the number of microphones which are used. Single-channel source separation methods typically exploit spectro-temporal diversity between the speech and the noise signals \cite{Kolbak_2017_IEEE_ACM_ASLP, Luo_2019_IEEE_ACM_ASLP, FurcaNet_2019, Wisdom_2019_ICASSP, WHAMR_ICASSP_2020, Luo_2020_ICASSP, Wang_2021_ICASSP, Cem_2021_ICASSP}. These methods typically perform source separation by estimating masks corresponding to each sound source using convolutional, recurrent or transformer networks. 
To improve the source separation performance, these methods also aim to learn short-term and long-term temporal dependencies of speech signals by neural structures, which have large receptive fields \cite{Luo_2019_IEEE_ACM_ASLP, FurcaNet_2019}, deep and wide recurrent layers \cite{Kolbak_2017_IEEE_ACM_ASLP, WHAMR_ICASSP_2020}, or dual path architectures using recurrent or transformer layers \cite{Luo_2020_ICASSP, Wang_2021_ICASSP, Cem_2021_ICASSP}.


When multiple microphones are available, multi-channel  filters allow to exploit the spatial diversity between the speakers and the background noise in addition to the spectro-temporal diversity \cite{Gannot2017, Ochiai_2020_ICASSP, Aroudi_2021_ICASSP, Tomohiro_2021_ICASSP, Dong_2021_ADL_MVDR, Wang_2021_Complex_Spectral_Mapping, Dong_Yu_2021_Interspeech_Generalized_Spatio_Temporal_RNN}. Multi-channel  filters, also often referred to as spatial filters and beamformers \cite{Simon2015, Gannot2017}, perform source separation by linearly filtering and summing the microphone signals. Conventional multi-channel  filters are typically estimated based on a linear optimization problem and require estimates of certain parameters, e.g., covariance matrices, direction-of-arrivals (DOAs) or steering vectors of sound sources \cite{Gannot2017, Simon2015}. These parameters can be estimated based on masks obtained by, e.g., single-channel neural-network-based source separation techniques \cite{Ochiai_2020_ICASSP, Tomohiro_2021_ICASSP, Dong_Yu_2021_Interspeech_Generalized_Spatio_Temporal_RNN}, or can be estimated directly from microphone signals using neural networks in an end-to-end fashion, as proposed in a DOA-driven beamforming network (DBNet) \cite{Aroudi_2021_ICASSP}. 

Instead of formulating the multi-channel filter as a linear optimization problem, it has been recently proposed to directly estimate multi-channel filters by a generalized recurrent beamformer (GRNN-BF) network \cite{Dong_Yu_2021_Interspeech_Generalized_Spatio_Temporal_RNN}, learning a non-linear optimization solution. 
GRNN-BF network is able to estimate multi-channel filters from microphone signals, but also relies on a camera input, which may not be available in many applications. Aiming at end-to-end source separation by directly estimating multi-channel filters from only microphone signals, we propose a transformer-recurrent-U network (TRUNet) in this paper. 
To draw valid conclusions on reverberant sound source separation, we evaluate the proposed network on a challenging and realistic reverberant dataset, generated from measured room impulse responses of an actual microphone array.

\setlength{\textfloatsep}{6pt}
\begin{figure}[t]
 \centering
  \centerline{\includegraphics[width=8cm]{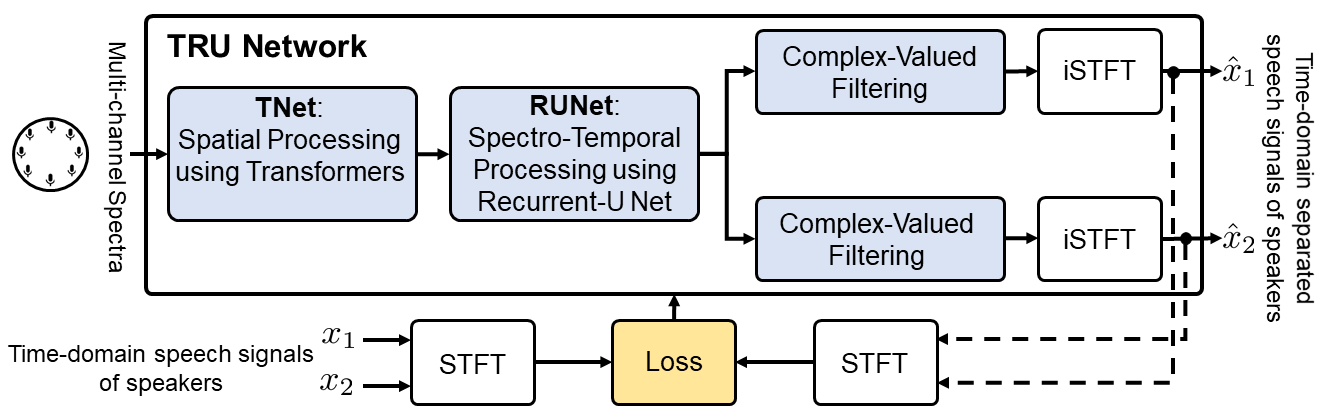}}
  \vspace{-0.35cm}
 \caption{\small Block diagram of the proposed TRUNet structure.}
\label{fig:TRUNet}
\end{figure}

The proposed TRUNet is depicted in Fig. \ref{fig:TRUNet}. TRUNet consists of a spatial processing unit using a transformer network (TNet), a spectro-temporal processing unit using a recurrent-U network \ac{RUNet}, and an \ac{iSTFT} layer. First, it accepts spectra of the multi-channel signals as input. Then, the multi-channel filters are estimated by the spatial processing unit and the spectro-temporal processing unit in an end-to-end fashion. 
For the spatial processing unit, since capturing the spatial diversity is not straightforward, we propose three transformer network architectures operating across microphone channels. For the spectro-temporal processing unit, we adopt a multi-channel \ac{RUNet} to efficiently capture spectral and temporal dependencies corresponding to each speaker. To separate out speakers the multi-channel filters, which are complex-valued and time-varying, are applied to the multi-channel input spectra. In addition to multi-channel filters, we also consider estimating single-channel filters that can still benefit from a multi-channel spectro-temporal designed filter. Finally, the separated sources are transformed to the time domain using the \ac{iSTFT} layer, which enforces \ac{STFT} consistency in the network \cite{Wisdom_2019_ICASSP, Braun_2021_ICASSP}.

We train the proposed network to separate out speakers from noisy and reverberant speech mixtures. The reverberant speech signals in the mixtures can be thought as containing early reflections, which are beneficial for speech intelligibility and naturalness, and a late reverberation component, which is known to have a detrimental effect on speech quality and intelligibility \cite{Warzybok_2013}. Therefore, we aim at separating out a low-reverberant speech signal preserving the early reflections. We use the complex mean-squared error (MSE) loss function compressed with an exponent factor to balance the optimization of small and large MSE errors \cite{Ephrat_2018, Braun_2021_ICASSP}, which was found to be superior to other losses for source separation \cite{Aroudi_2021_ICASSP}. We also explore the impact of the exponent factor on the source separation performance. 
Furthermore, we train the proposed network on a large dataset accounting for the crucial aspects of realistic multi-channel audio such as a large number of  speakers, various noise types, different microphone signal levels and reverberation as in \cite{Aroudi_2021_ICASSP, Braun_2021_ICASSP}.

\section{Sound source separation system}
\label{sec:Sound source separation network}

We consider an acoustic scenario comprising two competing speakers and background noise in a reverberant environment. TRUNet accepts spectra of the multi-channel signals as input. Aiming at separating out the speakers, the network estimates two complex-valued filters, using the spatial processing network and the spectro-temporal processing network. The multi-channel filtering on $M$-channel signals is performed as 
\begin{equation}
    \hat{X}_{i}\left(k,f\right)=\mathbf{B}_{i}\left (k,f \right)^{H}\mathbf{Y}\left(k,f \right),
\label{eq:multi-channel filtering }
\end{equation}
where $\hat{X}_{i}\left(k,f\right)$ denotes the separated speech signal corresponding to speaker $i$ in the \ac{STFT} domain, $\mathbf{B}_{i}\left(k,f \right)\in \mathbb{C}^{M\times 1}$ denotes the multi-channel filter directly estimated by the network, $\mathbf{Y}\left(k,f \right)\in \mathbb{C}^{M\times 1}$ denotes the stacked vector of all microphone signals, $\left ( \cdot  \right )^{H}$ denotes the conjugate transpose operator, and $k$ and $f$ are the frame index and the frequency index. 
In addition, a single-channel filtering version of (\ref{eq:multi-channel filtering }) is considered, i.e., 
$B_{i}\left (k,f \right)Y\left(k,f \right)$, where $B_{i}$ denotes the single-channel complex-valued filter and $Y\left(k,f \right)$ denotes one arbitrarily selected microphone signal in the STFT domain. 
In the following, we present the spatial processing unit and the spectro-temporal processing unit of the proposed TRUNet architecture.

\subsection{Spatial processing unit using TNet}
\label{subsec: Spatial process using transformer networks}
The proposed spatial processing unit  is a TNet, which consists of $N$ spatial transformer blocks and operates across microphone channels.   The spatial transformer blocks accept spectra as the input and have a spatial attention function with an output, a representation incorporating inter-channel information. For spatial transformer blocks we adopt transformers proposed for language translation tasks in \cite{attention_Google_2017}. The inputs of each transformer block consist of three pairs, i.e., queries $\mathbf{z}_{\mathbf{q}}\in \mathbb{R}^{M\times D}$ with $D$ the feature dimension, keys $\mathbf{z}_{\mathbf{k}}\in \mathbb{R}^{M\times D}$, and values $\mathbf{z}_{\mathbf{v}}\in \mathbb{R}^{M\times D}$, which can be real and imaginary parts or the magnitude and the phase of the input spectra. To direct the attention of a transformer block to sub-spaces of spectral feature space, the keys and values are linearly projected as 
\begin{equation}
     \mathbf{q}_{h}=\mathbf{z}_{\mathbf{q}}\mathbf{w}_{\mathbf{q},h}, \quad \mathbf{k}_{h}=\mathbf{z}_{\mathbf{k}}\mathbf{w}_{\mathbf{k},h}, \quad \mathbf{v}_{h}=\mathbf{z}_{\mathbf{v}}\mathbf{w}_{\mathbf{v},h},
    \label{eq:projection}
\end{equation}
where $h\in\left \{ 1 \; \cdots \;\;  \mathcal{H}\right \}$ denotes the sub-space index, also referred to as  heads \cite{attention_Google_2017}, and $\mathbf{w}_{\mathbf{q},h}, \mathbf{w}_{\mathbf{k},h}, \mathbf{w}_{\mathbf{v},h}\in \mathbb{R}^{D\times \frac{D}{\mathcal{H}}}$ are learnable projection matrices.  By this projection,  the spatial attention function is applied on the sub-spaces  in parallel with a head embedding dimension $\frac{D}{\mathcal{H}}$, speeding up the process. The spatial attention function is then performed by weighting the sum of the values, where the weights are computed by (real-valued) dot products of the queries and the keys,  followed by a softmax function, as
\begin{equation}
    \mathbf{a}_{h}=\textrm{Attention}\left ( \mathbf{q}_{h}, \; \mathbf{k}_{h}, \; \mathbf{v}_{h} \right )=\mathrm{softmax}\left ( \frac{\mathbf{q}_{h}\mathbf{k}_{h}^{T}}{\sqrt{D/\mathcal{H}}} \right )\mathbf{v}_{h},
    \label{eq:attention}
\end{equation}
where $\left ( \cdot  \right )^{T}$ denotes the transpose operator. The dot product operation in (\ref{eq:attention}) could be seen as a similar way as covariance matrices in conventional beamforming may be computed, and weighting the sum of the values could be seen as a similar way as beamforming weights may be computed.  
To allow a transformer block to jointly attend to information from different representation sub-spaces at different channels, the attention outputs of heads are concatenated and linearly projected with $\mathbf{w}_{\textrm{MH}}\in \mathbb{R}^{M\times D}$, using a multi-head attention \cite{attention_Google_2017}, i.e., $\mathbf{z}_{\textrm{MH}}=\left [ \mathbf{a}_{1} \; \cdots \;\; \mathbf{a}_{\mathcal{H}} \right ]\mathbf{w}_{\textrm{MH}}$. 
The multi-head attention output with a corresponding residual connection together are then followed by a layer normalization and a fully connected feed-forward network  \cite{attention_Google_2017}, resulting the output of a transformer block $\mathbf{z}_{\textrm{O}}$.

\setlength{\textfloatsep}{6pt}
\begin{figure}[t]
\begin{minipage}[b]{.49\linewidth}
  \centering
  \centerline{\includegraphics[width=2cm]{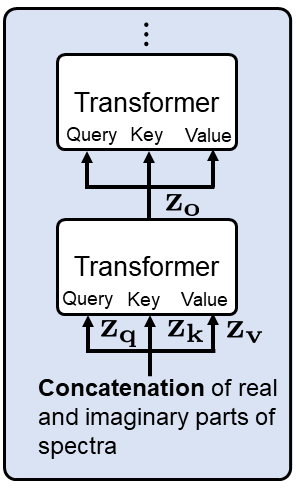}}
\centerline{\;\;\;(a) }\medskip
\end{minipage}%
\begin{minipage}[b]{.3\linewidth}
  \centering
  \centerline{\includegraphics[width=4cm]{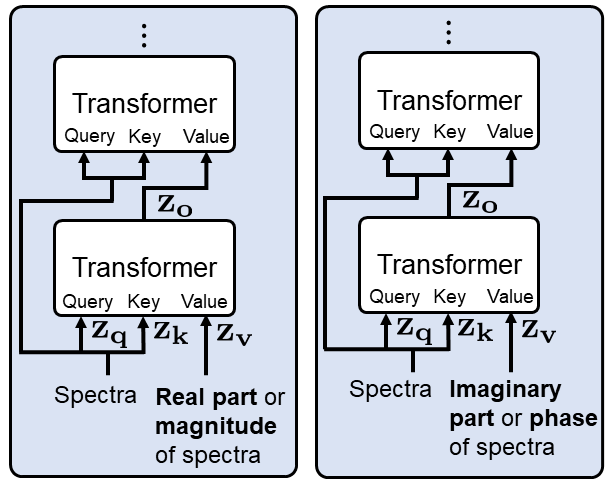}}
\centerline{\;(b) }\medskip
\end{minipage}
\vspace{-0.4cm}
\caption{\small Transformer blocks used in (a) TNet–Cat and (b) TNet–MagPhase and TNet–RealImag.}
\label{fig:Transformer}
\end{figure}

Since leveraging spatial diversity into a network is not straightforward, we consider several variants for computing cross-channel attention into the proposed transformer networks:
\begin{compactitem}
    \item \textbf{TNet--Cat} consists of several transformer blocks using the spatial attention function in (\ref{eq:attention}). The output of each transformer block $\mathbf{z}_{\textrm{O}}$ is used as the input to all query, key and value matrices of the next block, i.e. $\mathbf{z_q}=\mathbf{z_k}=\mathbf{z_v}$. The input of the first transformer block is the concatenation of the real and imaginary parts of the spectra (see Figure \ref{fig:Transformer}a). This approach simply computes the attention based on the real-valued dot product (\ref{eq:attention}) and can be seen as a straightforward approach to combining all multi-channel spectra into the proposed spatial transformers.

    \item \textbf{TNet--RealImag} uses two separate transformer stacks for real and imaginary parts, respectively. Queries and keys are all computed from the multi-channel spectra, i.e. $\mathbf{z}_{\mathbf{q}},\;\mathbf{z}_{\mathbf{k}}\in \mathbb{C}^{M\times D}$. Since the softmax function is not well-defined for complex arithmetic, the magnitude of the complex inner product is used instead, i.e. $\mathrm{softmax}\left ( \left | \frac{\mathbf{q}_{h}\mathbf{k}_{h}^{H}}{\sqrt{D/\mathcal{H}}} \right | \right )\mathbf{v}_{h}$, where $\left | \cdot  \right |$ denotes the magnitude operator and  $\left ( \cdot  \right )^{H}$ denotes the conjugate transpose operator. The output of the network is computed as the concatenation of the outputs of the last real and imaginary transformer stack blocks. As the real and imaginary parts are processed separately, this approach may still not be able to directly exploit the spatial information between channels, e.g., phase differences.

    \item \textbf{TNet--MagPhase} is analogous to TNet--RealImag, except that the spectral magnitude and the spectral phase are used instead of the real and imaginary parts. 
    The output of the network is computed as the concatenation of the outputs of the last spectral magnitude and phase transformer stack blocks.

\end{compactitem}

\subsection{Spectro-temporal processing unit using RUNet}
\label{subsec: Spectro-temporal processing using convolutional and recurrent networks}
Since the proposed TNets are not designed to capture spectral and temporal diversities in addition to the spatial diversity, the source separation capability of filters estimated by them is limited.
Therefore, we propose an end-to-end extension of TNets which incorporates a RUNet \cite{Tan_ICASS_2019, Aroudi_2021_ICASSP, Braun_2021_ICASSP}, aiming at also capturing spectral and temporal diversities.
In this work, we adopt a multi-channel \ac{RUNet}, accepting a multi-channel input which is a concatenation of the real and the imaginary part. It has $L$ symmetric convolutional and deconvolutional encoder and decoder layers with kernels of size $\left ( 6, \ 6 \right )$, aiming to deal with reverberation, and a stride of $\left ( 1, \ 2 \right )$ in time and frequency dimensions. The number of channels $C_{l}$ across layers $l\in \left \{ 1 \; \cdots \;\;  L\right \}$ increases per encoder layer, and decreases mirrored in the decoder. The input and the final output channels are $C_{\textrm{in}}=C_{\textrm{out}}=M$.  All convolutional layers are followed by leaky ReLU activations. The encoder and decoder are connected by  two BLSTM layers, which are fed with all features flattened along the channels. Aiming at a better network training while avoiding vanishing gradients, residual connections are used which link convolutional encoder layers and corresponding decoder layers. In addition, a residual connection linking the BLSTM layers was used.  Motivated by the results in \cite{Braun_2021_ICASSP} for speech enhancement, the residual connections are implemented as convolutions with $C_{l}$ channels and $\left ( 1, \ 1 \right )$ kernels. The network is then followed by a fully connected layer and a tanh activation to estimate multi-channel filters of both sources.  In addition to multi-channel filters, we consider to estimate single-channel filters using a similar network, however, with an extra convolutional layer after the last decoder layer with the output channel $C_{\textrm{out}}=1$. 
    
\subsection{Loss function}
\label{subsec:LOSS FUNCTION}
For training the networks we consider compressed MSE loss function (cMSE)  with an exponent factor $c$ \cite{Ephrat_2018, Wilson_2018_IWAENC} that balances small versus large amplitude contributions to MSE, i.e., 
\begin{equation}
    \mathcal{L}_{\mathrm{cMSE}}\left ( c \right ) = \log_{10}\underset{k,f}{\sum}\left| \left|X_{i}\left(k,f\right)\right|^{c}e^{j\varphi_{X}}-\left|\hat{X}_{i}\left(k,f\right)\right|^{c}e^{j\varphi_{\hat{X}}}\right| ^{2},
\end{equation}
where $\varphi_{X}$ and $\varphi_{\hat{X}}$ denote the spectral phase of the STFT of the target speech signal $X$ and separated speech signals, respectively.

To cope with the source-speaker to target-speaker mapping problem, we use utterance permutation invariant training (uPIT) \cite{Kolbak_2017_IEEE_ACM_ASLP}.


\section{Experimental setup}
\label{sec:Experimental setup}

\subsection{Dataset}
\label{subsec:Dataset}

We use realistic and large training and test sets to ensure generalization of our results to real-world acoustic conditions. 
For training, validation and testing, we use three different speech databases, i.e.\, 540~h of speech data from \cite{data_DNS-Challenge}, 40~h from VCTK \cite{VCTK_2}, and 5~h from DAPS \cite{DAPS}, respectively, and different noise databases from \cite{WHAM!, QUT_2010, Marc_dataset_2016}. We consider a 8-channel microphone array on a circle of 5~cm radius. 

For training, we simulate RIR sets of random positions in 1000 differently sized rooms using the image method \cite{Allen1979}, while for validation and testing we use measured RIRs using the actual microphone array in 10 different rooms. The rooms were with reverberation times between 0.2 to 0.8~s and direct to reverberant ratios between -12 to 5.8~dB.  

For data generation, we consider a similar data generation and augmentation pipeline as in \cite{Aroudi_2021_ICASSP, Braun_2021_ICASSP}: two overlapping speech signals of 30~s length are convolved with a RIR from a randomly chosen position in the same room and mixed with energy ratios drawn from a Gaussian distribution with $\mathcal{N}(0,\;2)$ dB. 
The reverberant mixture and noise are then mixed with a signal-to-noise ratio (SNR) drawn from a Gaussian distribution with $\mathcal{N}(8,\;10)$ dB. The resulting mixture signals are finally re-scaled to levels distributed with $\mathcal{N}(-28,\;10)$ dBFS. The target speech signals are generated as low-reverberant speech signals using the reverberant impulse responses shaped with a an exponential decay function \cite{Braun_2021_ICASSP}, enforcing a maximum reverberation time of 200~ms. We generate training, validation and test sets of 1000~h, 4~h and 4~h, respectively, at a sampling rate of $f_{s}=16$ kHz.

\subsection{Baseline method}
\label{subsec:Baseline method}
We consider a number of end-to-end \textit{multi-channel} source separation methods in our experiments: 1) \textbf{DBNet}, combining direction-of-arrival estimation and conventional spatial filtering \cite{Aroudi_2021_ICASSP}. 2) \textbf{eDBNet}, a DBnet extension using post-masking via convolutional-recurrent networks \cite{Aroudi_2021_ICASSP}. 3) \textbf{RUNet}, similar to the proposed TRUNet, but without a TNet spatial processing unit. 
We also consider a number of \textit{single-channel} methods: 1) \textbf{maskDaulPathTNet},  dual-path transformer masking-based network as in \cite{Cem_2021_ICASSP}. 2) \textbf{maskRUNet}, recurrent convolutional masking-based networks with U structure, as in \cite{Aroudi_2021_ICASSP}, but with residual connections as proposed in RUNet. 3) \textbf{maskDaulPathTRUNet}, consisting of maskDaulPathTNet followed by maskRUNet.

\subsection{Algorithmic parameters}
\label{subsec:  Algorithmic parameters}

For TNets and their extensions, we set the number of spatial transformer blocks\footnote{We also considered $N=6$ and $8$, but no significant performance improvement was observed.} $N=4$. In addition, all transformer blocks were used with positional encoding \cite{attention_Google_2017}. 
For the networks using  recurrent convolutional network with U structure, we use a sequence of 5 layers, i.e., $L=5$, with $16$, $16$, $32$, $32$ and $64$ filters. 
For the networks using recurrent layers, BLSTM layers with 1200 units are used. 
For maskDaulPathTNet, we use 4 layers of transformers in each path. 
We train all networks using the loss function $\mathcal{L}_{\mathrm{cMSE}}\left ( c=0.3 \right )$, except where explicitly different loss function is used. All networks were trained Adam optimizer \cite{Kingma2014}. In addition, we use gradient clipping technique with a maximum $L_{2}$ norm of 5, similarly as used in \cite{WHAMR_ICASSP_2020}.

\section{Experimental results}
\label{sec: Experimental results}
In this section, we evaluate the speech separation performance of the proposed networks in terms of the scale-invariant signal-to-distortion ratio (SDR) and the signal-to-noise ratio (SIR) of BSSEval [26] and PESQ [27]. In Section \ref{subsec: Spatial processing network source separation performance}, we investigate the source separation performance of the networks which incorporate only a TNet spatial processing unit. In Section \ref{subsec: TRUNet source separation performance}, we investigate the performance of the proposed TRUNet which incorporates both TNet spatial processing unit and RUNet spectro-temporal processing unit, and benchmark it against multi-channel and single-channel baseline methods. 

\subsection{TNet source separation performance}
\label{subsec: Spatial processing network source separation performance}

\begin{table}[t]
\renewcommand{\arraystretch}{.98}
\caption{\small Comparison of spatial transformer networks (TNets) using different number of attention heads $\mathcal{H}$ and embedding size $\frac{D}{\mathcal{H}}$.}
\vspace{-0.2cm}
\centering
\resizebox{\columnwidth}{!}{
\begin{tabular}{cccccc}
\hline
\bfseries Method & \bfseries Heads & \bfseries 
Embedding Size & \bfseries $\triangle\textrm{SDR}$ & \bfseries $\triangle\textrm{SIR}$ & \bfseries $\triangle\textrm{PESQ}$ \\
\hline\hline
DBNet\cite{Aroudi_2021_ICASSP} & - &  - &$5.65$ & $0.05$ & $0.01$\\ 
\hline
\multirow{3}{*}{TNet--Cat} 
    & $16$ & $64$ &$8.22$ & $1.14$ & $0.02$ \\
    & $64$ & $16$ & $8.22$ & $0.57$ & $0.02$\\ 
    & $256$ & $1$ & $8.23$ & $0.47$ & $0.01$\\ 
\hline
\multirow{3}{*}{TNet--RealImag}
    & $16$ & $64$ & $8.24$ & $1.83$ & $0.02$\\
    & $64$ & $16$ & $8.29$ & $0.92$ & $0.01$\\ 
    & $256$ & $1$ & $8.34$ & $0.71$ & $0.01$\\ 
\hline
\multirow{3}{*}{TNet--MagPhase}
    & $16$ & $64$ & $8.18$ & $\mathbf{2.55}$ & $0.02$\\
    & $64$ & $16$ & $8.22$ & $2.53$ & $0.01$\\ 
    & $256$ & $1$ & $8.36$ & $1.22$ & $0.01$\\ 
\hline
\end{tabular}
}
\label{tab:Tnet}
\end{table}

In Table \ref{tab:Tnet} the source separation performance of the networks incorporating a TNet spatial processing unit using different queries, keys and values and different number of attention heads $\mathcal{H}$ are compared with DBnet, which consists of a network-based DOA estimator and conventional spatial filters. We observe that TNets result in larger SDR and SIR improvements compared to DBnet, indicating that TNets are better able to spatial filter. The largest SIR improvement, indicating how well the speakers are separated, is obtained by \textbf{TNet--MagPhase}. In our opinion, this is due to the fact that queries and keys of TNet--MagPhase are computed from the complex-valued spectra, from which spatial information are straightforward to extract, and values of TNet--MagPhase are computed from the magnitude and the phase, which directly has the spatial information between channels.
Nevertheless, the improvement of TNets is limited particularly for the SIR of about $2.55$~dB, which can be mainly attributed to the limited capability of TNets to efficiently exploit spectral and temporal diversities in addition to the spatial diversity. We also observe that a lower number of attention heads consistently results in a larger SIR. A lower number of attention heads results in more global attention across spectrum sub-spaces as well as larger embedding size. 
We focus from now on systems using with $16$ heads, as they outperform the other settings.

We use an STFT frame length of 512 samples, an overlap of $50\%$ between successive frames, a Hann window and an FFT size $N_\text{FFT}=512$.

\subsection{TRUNet source separation performance}
\label{subsec: TRUNet source separation performance}

\begin{table}[t]
\renewcommand{\arraystretch}{.98}
\caption{\small Comparison of TRUNet with multi-channel baseline methods. \text{*} indicates methods accepting multi-channel inputs, but using single-channel complex-valued filtering.}
\vspace{-0.2cm}
\centering
\resizebox{\columnwidth}{!}{
\begin{tabular}{ccccc}
\hline
\bfseries Method & \bfseries $\triangle\textrm{SDR}$ & \bfseries $\triangle\textrm{SIR}$ & \bfseries $\triangle\textrm{PESQ}$ & \bfseries $\textrm{Model Size [M]}$\\
\hline
\hline
\multicolumn{5}{c}{\bfseries Multi-channel Baseline} \\ 
\hline
DBNet\cite{Aroudi_2021_ICASSP} &$5.65$ & $0.05$ & $0.01$ & $14$\\ 
eDBNet\cite{Aroudi_2021_ICASSP} & $8.50$ & $5.31$ & $0.20$ & $67$\\ 
RUNet & $8.47$ & $6.82$ & $0.14$ & $54$\\ 
RUNet\text{*} & $8.44$ & $7.35$ & $0.21$ & $54$\\ 
\hline
\multicolumn{5}{c}{\textbf{TNets}} \\ 
\hline
TNet--Cat & $8.22$ & $1.14$ & $0.02$ & $16$\\
TNet--RealImag & $8.24$ & $1.83$ & $0.02$ & $16$\\
TNet--MagPhase & $8.18$ & $2.55$ & $0.02$ & $16$\\
\hline
\multicolumn{5}{c}{\bfseries TRUNets} \\ 
\hline
TRUNet--Cat & $8.87$ & $6.98$ & $0.13$ & $31$\\ 
TRUNet--Cat\text{*} & $9.05$ & $12.51$ & $0.22$ & $31$\\
TRUNet--RealImag\text{*} & $9.12$ & $11.08$ & $0.15$ & $29$\\
TRUNet--MagPhase\text{*} & $\mathbf{9.38}$ & $\mathbf{12.87}$ & $\mathbf{0.22}$ & $29$\\ 
\hline
\end{tabular}
\label{tab:multi-ch}
\bfseries }
\end{table}

\begin{table}[t]
\renewcommand{\arraystretch}{.98}
\caption{\small Comparison of TRUNet--MagPhase\text{*} with single-channel baseline methods.}
\vspace{-0.2cm}
\centering
\resizebox{\columnwidth}{!}{
\begin{tabular}{ccccc}
\hline
\bfseries Method & \bfseries $\triangle\textrm{SDR}$ & \bfseries $\triangle\textrm{SIR}$ & \bfseries $\triangle\textrm{PESQ}$ & \bfseries $\textrm{Model Size [M]}$\\
\hline\hline
maskRUNet\cite{Aroudi_2021_ICASSP} & $9.57$ & $9.05$ & $0.07$ & $52$\\
maskDaulPathTNet\cite{Cem_2021_ICASSP} & $9.36$ & $10.36$ & $0.10$ & $11$\\
maskDaulPathTRUNet & $\mathbf{9.81}$ & $10.16$ & $0.10$ & $18$ \\ 
\hline
TRUNet--MagPhase\text{*} & $9.38$ & $\mathbf{12.87}$ & $\mathbf{0.22}$ & $29$\\ 
\hline
\end{tabular}
\label{tab:single-ch}
}
\end{table}

Table \ref{tab:multi-ch} shows the source separation performance of TRUNets, incorporating a TNet spatial processing
unit and a RUNet spectro-temporal processing unit,  versus the multi-channel baseline methods and the networks incorporating only TNets. Please note that TRUNets with {*} indicate the methods accepting multi-channel inputs, but using single-channel complex-valued filtering.
We observe that TRUNets result in larger performance measures, particularly for the SIR and the PESQ improvements, compared to TNets, indicating the importance of both spatial processing and spectro-temporal processing units for source separation performance. We also observe that only some TRUNets (TRUNet--Cat\text{*}, TRUNet--RealImag\text{*}, TRUNet--MagPhase\text{*}) result in a larger SDR improvement (about $9.0-9.3$~dB) and a larger SIR improvement (about $11.08-12.8$~dB) compared to the multi-channel baseline methods. Therefore, we investigate the main factors contributing to the performance measure improvement of TRUNets in the remainder.

We observe that TRUNet--Cat\text{*}, which uses single-channel filtering, compared to TRUNet-Cat, which uses multi-channel filtering, with a gain of $5.5$~dB for the SIR improvement and a gain of 0.09 for the PESQ improvement. This may imply that, for the considered networks, single-channel filtering is sufficient and summing all microphone spectra after filtering may be unnecessary. Among the TRUNets with single-channel filtering, the network using the spectral phase and the spectral magnitude (TRUNet--MagPhase\text{*}) obtains the best SDR improvement of $\textbf{9.38}$ and SIR improvement of $\textbf{12.87}$. In addition, TRUNet--MagPhase\text{*} yields the largest performance measures even compared to all other multi-channel methods. 

We further compare the source separation performance of TRUNet--MagPhase\text{*} with the single-channel baseline methods (see Table \ref{tab:single-ch}). 
We observe that although all considered methods result in a similar SDR improvement of about $9.60-9.81$ dB, the proposed TRUNet--MagPhase\text{*} stands out with a larger SIR improvement of $12.87$ and a PESQ improvement of $0.22$. 
The better SIR improvement of  TRUNet-MagPhase method indicates that the proposed method is able to better separate speakers while the similar SDR improvement might imply that the method may not be competitive in terms of ISR and SAR, which have not been investigated in this paper.

\begin{figure}[t]
 \centering
  \centerline{\includegraphics[width=8.9cm]{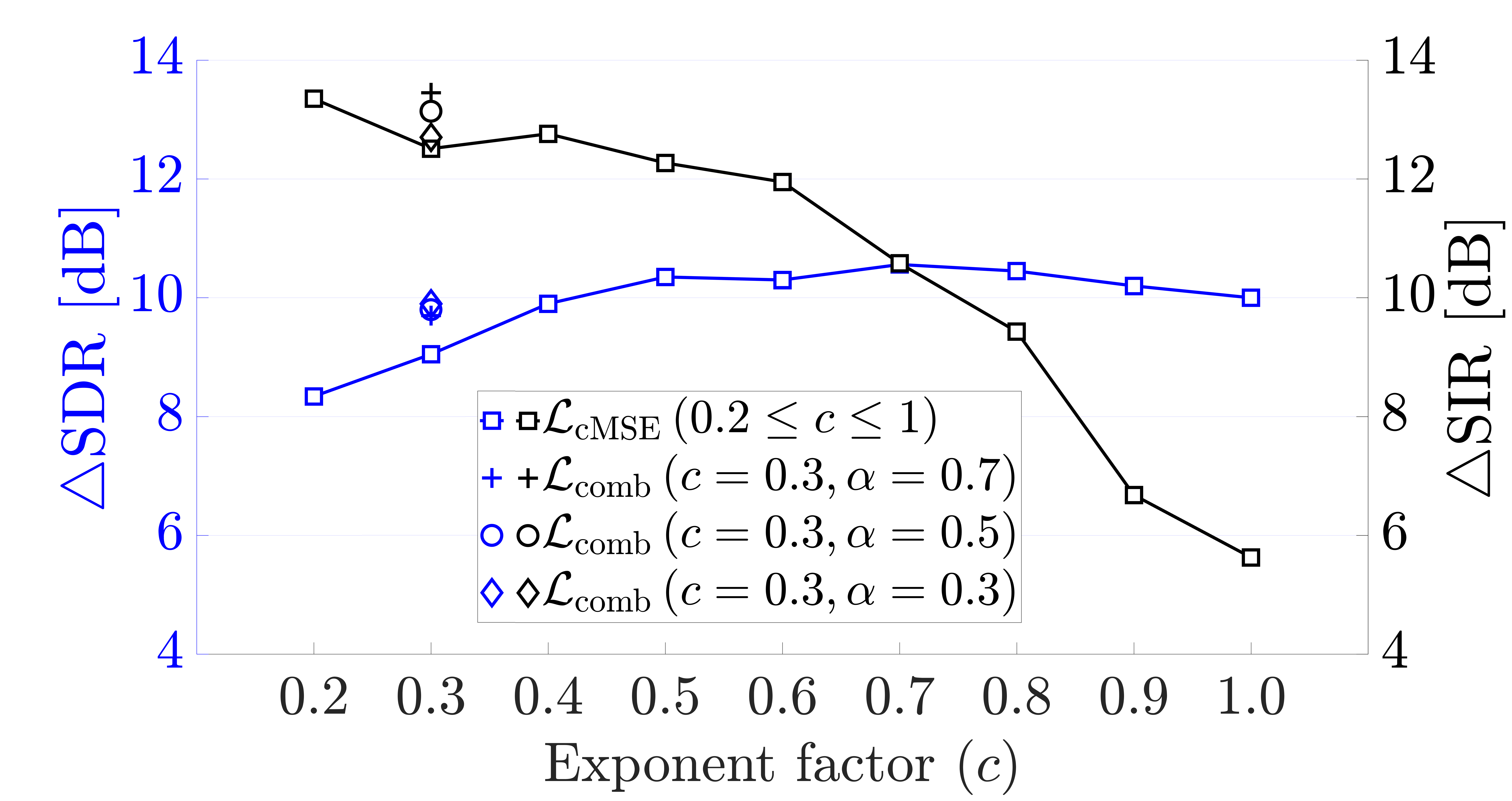}}
  \vspace{-0.3cm}
 \caption{\small SDR and SIR improvement for the compressed loss function  $\mathcal{L}_{\mathrm{cMSE}}\left ( c \right )$\protect\footnotemark and $\mathcal{L}_{\mathrm{comb}}\left ( c,\;\;\alpha \right )$ when using TRUNet--MagPhase\text{*}. The blue data points represent the SDR improvement and the black data points represent the SIR improvement.}
\label{fig:SDR_SIR}
\end{figure}

Finally, we explore the impact of the exponent factor on the source separation performance of TRUNet--MagPhase\text{*}. Figure \ref{fig:SDR_SIR} depicts SDR and SIR improvements, when the compressed loss function $\mathcal{L}_{\mathrm{cMSE}}\left ( c \right )$ is used. Smaller exponent factors are shown to obtain a larger SIR improvement (about $5.6$ to $13.3$~dB) and a smaller SDR improvement (about $10.5$ to $8.3$~db), compared to large factors. In order to achieve large improvements for both SDR and SIR, we finally experiment with linearly combining two cMSE losses with complementary exponent factors, i.e. $\mathcal{L}_{\mathrm{comb}}\left ( c,\;\;\alpha \right ) =\alpha\mathcal{L}_{\mathrm{cMSE}}\left ( c \right )+\left(1-\alpha\right)\mathcal{L}_{\mathrm{cMSE}}\left ( 1-c \right )$. When combining the compressed loss functions using $c=0.3$ and the combination factor $\alpha=0.7$, we obtain additional improvements resulting in final SDR and SIR of $\textbf{9.70}$ and $\textbf{13.40}$, respectively.

\footnotetext{$\mathcal{L}_{\mathrm{cMSE}}\left ( c \right )$ with $c= 0.1$ results in an unstable network training.}

\section{Conclusion}
\label{sec: Conclusion}
We proposed an end-to-end multi-channel source separation network that directly estimates multi-channel filters from multi-channel input spectra. The network consists of a spatial processing unit using transformers and a spectro-temporal processing unit using a recurrent U-structured convolutional network. In addition to multi-channel filters, we also consider estimating single-channel filters from multi-channel input spectra using TRUNet. We trained the network using a proposed combined cMSE loss function on a large reverberant dataset, and tested on realistic data using measured RIRs from an actual microphone array. The experimental results show that both proposed spatial and spectro-temporal processing units are crucial to obtain competitive performance. In particular, the results show that the proposed transformer-based spatial processing unit is better able to perform spatial filtering compared to networks using conventional spatial filtering. 
Moreover, the results show that our proposed architecture TRUNet achieves larger separation performance with single-channel filtering than multi-channel filtering, even larger than the performance obtained by the state-of-the-art source separation methods. 


\balance

\bibliographystyle{IEEEtran}
\bibliography{mainRefs_2_2_compact}

\begin{thebibliography}{10}
\providecommand{\url}[1]{#1}
\csname url@samestyle\endcsname
\providecommand{\newblock}{\relax}
\providecommand{\bibinfo}[2]{#2}
\providecommand{\BIBentrySTDinterwordspacing}{\spaceskip=0pt\relax}
\providecommand{\BIBentryALTinterwordstretchfactor}{4}
\providecommand{\BIBentryALTinterwordspacing}{\spaceskip=\fontdimen2\font plus
\BIBentryALTinterwordstretchfactor\fontdimen3\font minus
  \fontdimen4\font\relax}
\providecommand{\BIBforeignlanguage}[2]{{%
\expandafter\ifx\csname l@#1\endcsname\relax
\typeout{** WARNING: IEEEtran.bst: No hyphenation pattern has been}%
\typeout{** loaded for the language `#1'. Using the pattern for}%
\typeout{** the default language instead.}%
\else
\language=\csname l@#1\endcsname
\fi
#2}}
\providecommand{\BIBdecl}{\relax}
\BIBdecl

\bibitem{Kolbak_2017_IEEE_ACM_ASLP}
M.~{Kolb{\ae}k}, D.~{Yu}, Z.~{Tan}, and J.~{Jensen}, ``Multi-talker speech
  separation with utterance-level permutation invariant training of deep
  recurrent neural networks,'' \emph{IEEE/ACM Trans. Audio, Speech, Language
  Process.}, vol.~25, no.~10, pp. 1901--1913, Oct. 2017.

\bibitem{Luo_2019_IEEE_ACM_ASLP}
Y.~{Luo} and N.~{Mesgarani}, ``Conv-tasnet: Surpassing ideal time–frequency
  magnitude masking for speech separation,'' \emph{IEEE/ACM Trans. Audio,
  Speech, Language Process.}, vol.~27, no.~8, pp. 1256--1266, Aug. 2019.

\bibitem{FurcaNet_2019}
Z.~Shi, H.~Lin, L.~Liu, R.~Liu, S.~Hayakawa, S.~Harada, and J.~Han,
  ``{FurcaNet}: An end-to-end deep gated convolutional, long short-term memory,
  deep neural networks for single channel speech separation,'' arXiv preprint,
  2019.

\bibitem{Wisdom_2019_ICASSP}
S.~Wisdom, J.~R. Hershey, K.~Wilson, J.~Thorpe, M.~Chinen, B.~Patton, and R.~A.
  Saurous, ``Differentiable consistency constraints for improved deep speech
  enhancement,'' in \emph{{ICASSP}}, May 2019, pp. 900--904.

\bibitem{WHAMR_ICASSP_2020}
M.~{Maciejewski}, G.~{Wichern}, E.~{McQuinn}, and J.~L. {Roux}, ``{WHAMR!}:
  Noisy and reverberant single-channel speech separation,'' in \emph{{ICASSP}},
  2020, pp. 696--700.

\bibitem{Luo_2020_ICASSP}
Y.~Luo, Z.~Chen, and T.~Yoshioka, ``Dual-path {RNN}: Efficient long sequence
  modeling for time-domain single-channel speech separation,'' in
  \emph{{ICASSP}}, May 2020, pp. 46--50.

\bibitem{Wang_2021_ICASSP}
K.~Wang, B.~He, and W.-P. Zhu, ``Tstnn: Two-stage transformer based neural
  network for speech enhancement in the time domain,'' in \emph{{ICASSP}}, June
  2021, pp. 7098--7102.

\bibitem{Cem_2021_ICASSP}
C.~Subakan, M.~Ravanelli, S.~Cornell, M.~Bronzi, and J.~Zhong, ``Attention is
  all you need in speech separation,'' in \emph{{ICASSP}}, Jun. 2021, pp.
  21--25.

\bibitem{Gannot2017}
S.~Gannot, E.~Vincent, S.~Markovich-Golan, and A.~Ozerov, ``A consolidated
  perspective on multimicrophone speech enhancement and source separation,''
  \emph{IEEE/ACM Trans. Audio, Speech, Language Process.}, vol.~25, no.~4, pp.
  692--730, 2017.

\bibitem{Ochiai_2020_ICASSP}
T.~{Ochiai}, M.~{Delcroix}, R.~{Ikeshita}, K.~{Kinoshita}, T.~{Nakatani}, and
  S.~{Araki}, ``{Beam-TasNet}: Time-domain audio separation network meets
  frequency-domain beamformer,'' in \emph{{ICASSP}}, 2020, pp. 6384--6388.

\bibitem{Aroudi_2021_ICASSP}
A.~Aroudi and S.~Braun, ``{DBNet}: {DOA}-driven beamforming network for
  end-to-end reverberant sound source separation,'' in \emph{{ICASSP}}, June
  2021, pp. 211--215.

\bibitem{Tomohiro_2021_ICASSP}
T.~Nakatani, R.~Ikeshita, K.~Kinoshita, H.~Sawada, and S.~Araki, ``Blind and
  neural network-guided convolutional beamformer for joint denoising,
  dereverberation, and source separation,'' in \emph{{ICASSP}}, Jun. 2021, pp.
  6129--6133.

\bibitem{Dong_2021_ADL_MVDR}
Z.~Zhang, Y.~Xu, M.~Yu, S.-X. Zhang, L.~Chen, D.~S. Williamson, and D.~Yu,
  ``Multi-channel multi-frame {ADL-MVDR} for target speech separation,''
  \emph{IEEE/ACM Trans. Audio, Speech, Language Process.}, vol.~29, pp.
  3526--3540, Nov. 2021.

\bibitem{Wang_2021_Complex_Spectral_Mapping}
Z.-Q. Wang, P.~Wang, and D.~Wang, ``Multi-microphone complex spectral mapping
  for utterance-wise and continuous speech separation,'' \emph{IEEE/ACM Trans.
  Audio, Speech, Language Process.}, vol.~29, pp. 2001--2014, May 2021.

\bibitem{Dong_Yu_2021_Interspeech_Generalized_Spatio_Temporal_RNN}
Y.~Xu, Z.~Zhang, and D.~Y. Meng~Yu, Shi-Xiong~Zhang, ``Generalized
  spatio-temporal rnn beamformer for target speech separation,'' in
  \emph{{INTERSPEECH}}, Brno, Czechia, Sep. 2021, pp. 3076--3080.

\bibitem{Simon2015}
S.~Doclo, W.~Kellermann, S.~Makino, and S.~E. Nordholm, ``Multichannel signal
  enhancement algorithms for assisted listening devices,'' \emph{IEEE Signal
  Process. Magazine}, vol.~32, no.~2, pp. 18--30, Mar. 2015.

\bibitem{Braun_2021_ICASSP}
S.~Braun, H.~Gamper, C.~K. Reddy, and I.~Tashev, ``Towards efficient models for
  real-time deep noise suppression,'' in \emph{{ICASSP}}, June 2021, pp.
  656--660.

\bibitem{Warzybok_2013}
A.~Warzybok, J.~Rennies, T.~Brand, S.~Doclo, and B.~Kollmeier, ``Effects of
  spatial and temporal integration of a single early reflection on speech
  intelligibility,'' \emph{The Journal of the Acoustical Society of America},
  vol. 133, no.~1, pp. 269--282, Jan 2013.

\bibitem{Ephrat_2018}
A.~Ephrat, I.~Mosseri, O.~Lang, T.~Dekel, K.~Wilson, A.~Hassidim, W.~T.
  Freeman, and M.~Rubinstein, ``Looking to listen at the cocktail party: A
  speaker-independent audio-visual model for speech separation,'' \emph{ACM
  Trans. Graph.}, vol.~37, no.~4, Jul. 2018.

\bibitem{attention_Google_2017}
A.~Vaswani, N.~Shazeer, N.~Parmar, J.~Uszkoreit, L.~Jones, A.~N. Gomez,
  L.~Kaiser, and I.~Polosukhin, ``Attention is all you need,'' in
  \emph{{NIPS}}.\hskip 1em plus 0.5em minus 0.4em\relax Red Hook, NY, USA:
  Curran Associates Inc., 2017, p. 6000–6010.

\bibitem{Tan_ICASS_2019}
K.~{Tan}, X.~{Zhang}, and D.~{Wang}, ``Real-time speech enhancement using an
  efficient convolutional recurrent network for dual-microphone mobile phones
  in close-talk scenarios,'' in \emph{{ICASSP}}, 2019, pp. 5751--5755.

\bibitem{Wilson_2018_IWAENC}
K.~{Wilson}, M.~{Chinen}, J.~{Thorpe}, B.~{Patton}, J.~{Hershey}, R.~A.
  {Saurous}, J.~{Skoglund}, and R.~F. {Lyon}, ``Exploring tradeoffs in models
  for low-latency speech enhancement,'' in \emph{{IWAENC}}, 2018, pp. 366--370.

\bibitem{data_DNS-Challenge}
``{IEEE ICASSP 2021 Deep Noise Suppression (DNS) Challenge},''
  \url{https://github.com/microsoft/DNS-Challenge}.

\bibitem{VCTK_2}
J.~Yamagishi, C.~Veaux, and K.~MacDonald, ``{CSTR VCTK} corpus: English
  multi-speaker corpus for {CSTR} voice cloning toolkit (version 0.92),
  [sound],'' in \emph{University of Edinburgh. The Centre for Speech Technology
  Research}, 2019.

\bibitem{DAPS}
``{Device and produced speech (DAPS) dataset},''
  \url{https://ccrma.stanford.edu/~gautham/Site/daps.html}.

\bibitem{WHAM!}
G.~Wichern, J.~Antognini, M.~Flynn, L.~R. Zhu, E.~McQuinn, D.~Crow, .~Manilow,
  and J.~Le~Roux, ``{WHAM!}: Extending speech separation to noisy
  environments,'' in \emph{{INTERSPEECH}}, Sep. 2019.

\bibitem{QUT_2010}
D.~Dean, S.~Sridharan, R.~Vogt, and M.~Mason, ``The qut-noise-timit corpus for
  evaluation of voice activity detection algorithms,'' in \emph{Proceedings of
  the Annual Conference of the International Speech Communication Association},
  2010, pp. 3110--3113.

\bibitem{Marc_dataset_2016}
M.~Ferras, S.~R. Madikeri, P.~Motl{\'{\i}}cek, S.~Dey, and H.~Bourlard, ``A
  large-scale open-source acoustic simulator for speaker recognition,''
  \emph{IEEE Signal Process. Lett.}, vol.~23, no.~4, pp. 527--531, 2016.

\bibitem{Allen1979}
J.~B. Allen and D.~A. Berkley, ``Image method for efficiently simulating
  small-room acoustics,'' \emph{The Journal of the Acoustical Society of
  America}, vol.~65, no.~4, pp. 943--950, Apr. 1979.

\bibitem{Kingma2014}
D.~Kingma and J.~Ba, ``Adam: A method for stochastic optimization,'' arXiv
  preprint, 2014.

\end{thebibliography}

\end{document}